\documentclass[twocolumn,10pt]{article}
\pdfoutput=1
\usepackage{times}
\usepackage{graphicx}
\usepackage{amsmath}
\usepackage{amssymb}
\usepackage{color}
\usepackage[cm]{fullpage}
\usepackage{multirow}
\usepackage{booktabs}
\usepackage{pbox}

\usepackage[sort,square]{natbib}
\newcommand{\bibfont}{\small}
\renewcommand{\cite}[1]{\citep{#1}}

\usepackage{authblk}

\usepackage[footnotesize,bf,sf,font={sf,footnotesize}]{caption}
\usepackage[rightcaption]{sidecap}

\usepackage{sectsty}
\sectionfont{\large}
\subsectionfont{\normalsize}

\usepackage{titlesec}
\titlespacing\section{0pt}{12pt plus 4pt minus 2pt}{0pt}
\titlespacing\subsection{0pt}{12pt plus 4pt minus 2pt}{0pt}
\titlespacing\subsubsection{0pt}{12pt plus 4pt minus 2pt}{0pt}

\title{\vspace{-1.0cm}Predicting the outcomes of treatment to eradicate the latent reservoir for HIV-1 }

\author[1,2,$\ast$]{Alison L. Hill}

\author[1,3,$\ast$]{Daniel I. S. Rosenbloom}
\author[4]{Feng Fu}
\author[1]{Martin A. Nowak}
\author[5,$\dagger$]{Robert F. Siliciano}

\affil[1]{Program for Evolutionary Dynamics, Department of Mathematics,Department of Organismic and Evolutionary Biology, Harvard University, Cambridge, MA 02138, USA}
\affil[2]{Biophysics Program and Harvard-MIT Division of Health Sciences and Technology, Harvard University, Cambridge, MA 02138, USA}
\affil[3]{Department of Biomedical Informatics, Columbia University Medical Center, New York, NY 10032, USA}
\affil[4]{Institute of Integrative Biology, ETH Zurich, 8092 Zurich, Switzerland}
\affil[5]{Department of Medicine, Johns Hopkins University School of Medicine and Howard Hughes Medical Institute, Baltimore, MD 21205, USA}
\affil[$\ast$]{These authors contributed equally to the manuscript}
\affil[$\dagger$]{To whom correspondence should be addressed: rsiliciano@jhmi.edu}

\date{}

\newcommand\cdfour{$\text{CD4}^{+}~$}

\newcommand\inv{$^{-1}$}
 \providecommand{\e}[1]{\ensuremath{\times 10^{#1}}}

\makeatletter
\DeclareRobustCommand{\text}{%
  \ifmmode\expandafter\text@\else\expandafter\mbox\fi}
\let\nfss@text\text
\def\text@#1{{\mathchoice
  {\textdef@\displaystyle\f@size{#1}}%
  {\textdef@\textstyle\f@size{#1}}%
  {\textdef@\textstyle\sf@size{#1}}%
  {\textdef@\textstyle \ssf@size{#1}}%
  \check@mathfonts
  }%
}
\def\textdef@#1#2#3{\hbox{{%
                    \everymath{#1}%
                    \let\f@size#2\selectfont
                    #3}}}
\makeatother

\begin{document}

\maketitle

\section*{Abstract}
{Massive research efforts are now underway to develop a cure for HIV infection, allowing patients to discontinue lifelong combination antiretroviral therapy (ART). New latency-reversing agents (LRAs) may be able to purge the persistent reservoir of latent virus in resting memory \cdfour T cells, but the degree of reservoir reduction needed for cure remains unknown. Here we use a stochastic model of infection dynamics to estimate the efficacy of LRA needed to prevent viral rebound after ART interruption. We incorporate clinical data to estimate population-level parameter distributions and outcomes. Our findings suggest that approximately 2{,}000-fold reductions are required to permit a majority of patients to interrupt ART for one year without rebound and that rebound may occur suddenly after multiple years. Greater than 10{,}000-fold reductions may be required to prevent rebound altogether. Our results predict large variation in rebound times following LRA therapy, which will complicate clinical management. This model provides benchmarks for moving LRAs from the lab to the clinic and can aid in the design and interpretation of clinical trials. These results also apply to other interventions to reduce the latent reservoir and can explain the observed return of viremia after months of apparent cure in recent bone marrow transplant recipients and an immediately-treated neonate. }

\section*{Significance Statement}
HIV infection cannot be cured by current antiretroviral drugs, due to the presence of long-lived latently-infected cells. New anti-latency drugs are being tested in clinical trials, but major unknowns remain. It is unclear how much latent virus must be eliminated for a cure, which remains difficult to answer empirically due to few case studies and limited sensitivity of viral reservoir assays.  In this paper, we introduce a mathematical model of HIV dynamics to calculate the likelihood and timing of viral rebound following anti-latency treatment. We derive predictions for the required efficacy of anti-latency drugs, and demonstrate that rebound times may be highly variable and occur after years of remission. These results will aid in designing and interpreting HIV cure studies.

\section*{Introduction}

The latent reservoir (LR) for HIV-1 is a population of long-lived resting memory \cdfour T cells with integrated HIV-1 DNA~\cite{chun_quantification_1997}. After establishment during acute infection~\cite{chun_early_1998}, it increases to $10^5 - 10^7$ cells and then remains stable. As only replicating virus is targeted by antiretroviral therapy (ART), latently infected cells persist even after years of effective treatment~\cite{siliciano_long-term_2003, archin_measuring_2014}. Cellular activation leads to virus production and, if treatment is interrupted, viremia rebounds within weeks~\cite{ruiz_structured_2000}. Several molecular mechanisms maintain latency, including epigenetic modifications, transcriptional interference from host genes, and the absence of activated transcription factors~\cite{marsden_establishment_2010, hakre_epigenetic_2011,mbonye_control_2011,ruelas_integrated_2013}. 

Major efforts are underway to identify pharmacologic agents that reverse latency by triggering the expression of HIV-1 genes in latently infected cells, with the hope that cell death from viral cytopathic effects or cytolytic immune responses follows, reducing the size of the LR~\cite{choudhary_curing_2011,durand_developing_2012}. Collectively called \emph{latency-reversing agents} (LRAs), these drugs include histone deacetylase inhibitors~\cite{archin_antiretroviral_2010,archin_administration_2012, shirakawa_reactivation_2013}, protein kinase C activators~\cite{korin_effects_2002,williams_prostratin_2004,mehla_bryostatin_2010,dechristopher_designed_2012}, and the bromodomain inhibitor JQ1~\cite{bartholomeeusen_bromodomain_2012,zhu_reactivation_2012,boehm_bet_2013}. While LRAs are the subject of intense research, it is unclear how much the LR must be reduced to enable patients to safely discontinue ART.

The feasibility of reservoir reduction as a method of HIV-1 cure is supported by case studies of stem-cell transplantation~\cite{hutter_long-term_2009, henrich_antiretroviral-free_2014} and, more recently, early treatment initiation~\cite{saez-cirion_post-treatment_2013,persaud_absence_2013}, which have allowed patients to interrupt treatment for months or years without viral rebound.  The dramatic reductions in reservoir size accompanying these strategies stands in stark contrast to the actions of current LRAs, which induce only a fraction of latent virus \emph{in vitro}~\cite{cillo_quantification_2014,bullen_new_2014} and have not produced a measurable decrease in LR size $\emph{in vivo}$~\cite{archin_antiretroviral_2010,archin_administration_2012,spivak_pilot_2013}. It unclear how patient outcomes depend on reservoir reduction between these extremes, nor even whether a reduction that falls short of those achieved with stem-cell transplantation will bring any clinical benefit. LRA research needs to address the question: \emph{how low must we go}? 

\begin{figure}[htb!]
\begin{center}
\includegraphics[width=0.8\columnwidth]{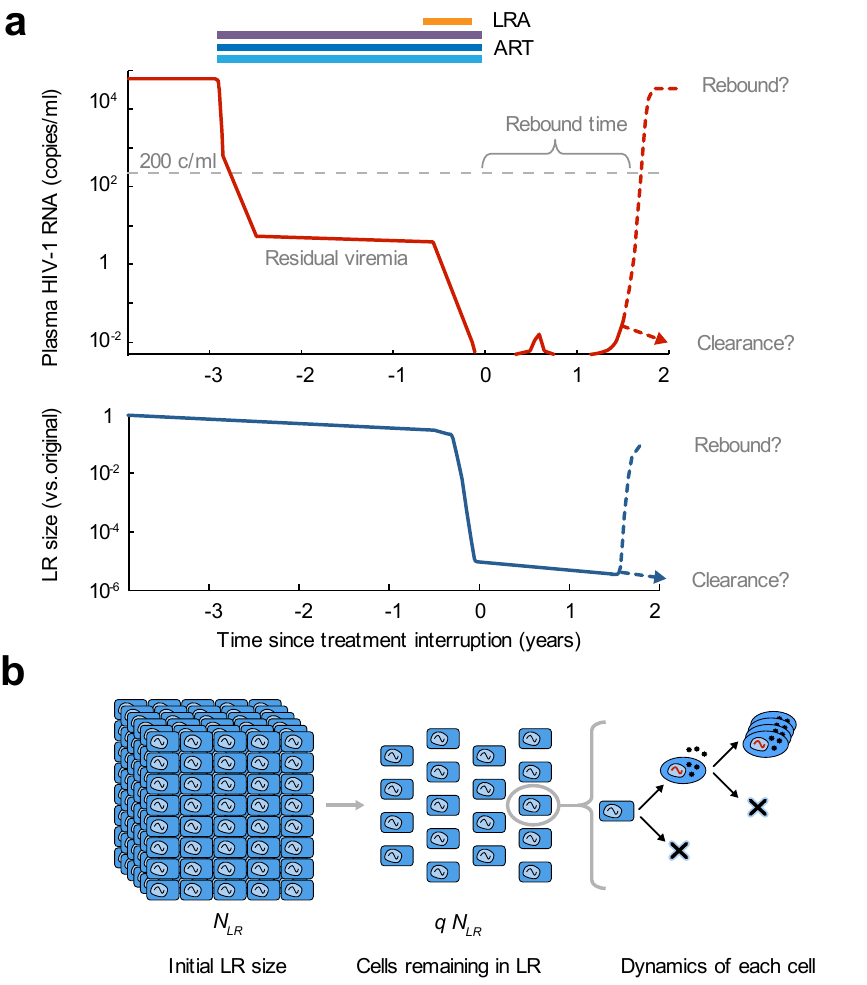}
\caption{Schematic of LRA treatment and stochastic model of rebound following interruption of ART. a) Proposed treatment protocol, illustrating possible viral load and size of LR before and after LRA therapy. When ART is started, viral load decreases rapidly and may fall below the limit of detection. The LR is established early in infection (not shown) and decays very slowly over time. When LRA is administered, the LR declines. After discontinuation of ART, the infection may be cleared, or viremia may eventually rebound. b) LRA efficacy is defined by the parameter $q$, the fraction of the LR remaining after therapy, which determines the initial conditions of the model. The stochastic model of viral dynamics following interruption of ART and LRA tracks both latently infected resting \cdfour T cells (rectangles) and productively infected \cdfour T cells (ovals). Each arrow represents an event that occurs in the model. Alternate models considering homeostatic proliferation and turnover of the LR are discussed in the Methods and Supplementary Methods. Viral rebound occurs if at least one remaining cell survives long enough to activate and produce a chain of infection events leading to detectable infection (plasma HIV-1 RNA $>$ 200 c ml$^{-1}$).}
\label{fig:modeldiagram}
\end{center}
\end{figure}

In the absence of clinical data, mechanistic mathematical models can serve as a framework to predict results of novel interventions and plan clinical trials. When results do become available, the models can be tested and refined. Mathematical models have a long tradition of informing HIV-1 research and have been particularly useful in understanding HIV-1 treatment. Previous models have explained the multi-phasic decay of viremia during antiretroviral therapy~\cite{perelson_decay_1997}, the initial seeding of the LR during acute infection~\cite{archin_immediate_2012}, the limited inflow to the LR during treatment~\cite{sedaghat_limits_2007}, the dynamics of viral blips~\cite{Conway:2011tf}, and the contributions of the LR to drug resistance~\cite{rosenbloom_antiretroviral_2012}. No model has yet been offered to describe the effect of LRAs. Here we present a novel modeling framework to predict the degree of reservoir reduction needed to prevent viral rebound following ART interruption. The model can be used to estimate the probability that cure is achieved, or, barring that outcome, to estimate the length of time following treatment interruption before viral rebound occurs (Fig.~\ref{fig:modeldiagram}a).

\section*{Results}
\subsection*{Determination of key viral dynamic parameters governing patient outcomes}
We employ a stochastic model of HIV-1 reservoir dynamics and rebound that, in its simplest form, tracks two cell types: productively infected activated \cdfour T cells and latently infected resting \cdfour T cells (Fig.~\ref{fig:modeldiagram}b). A latently infected cell can either activate or die, each with a particular rate constant. An actively infected cell can produce virions, resulting in the active infection of some number of other cells, or it can die from other causes without producing virions that infect other cells; in the latter case, cytotyoxic T lymphocyte (CTL) killing, errors in viral reverse transcription, or other problems upstream of virion production may prevent further infection. The model only tracks the initial stages of viral rebound, when target cells are not yet limited. A full description is provided in the Methods and Supplementary Methods. 

\begin{figure}[hbt]
\begin{center}
 \includegraphics[width=1\columnwidth]{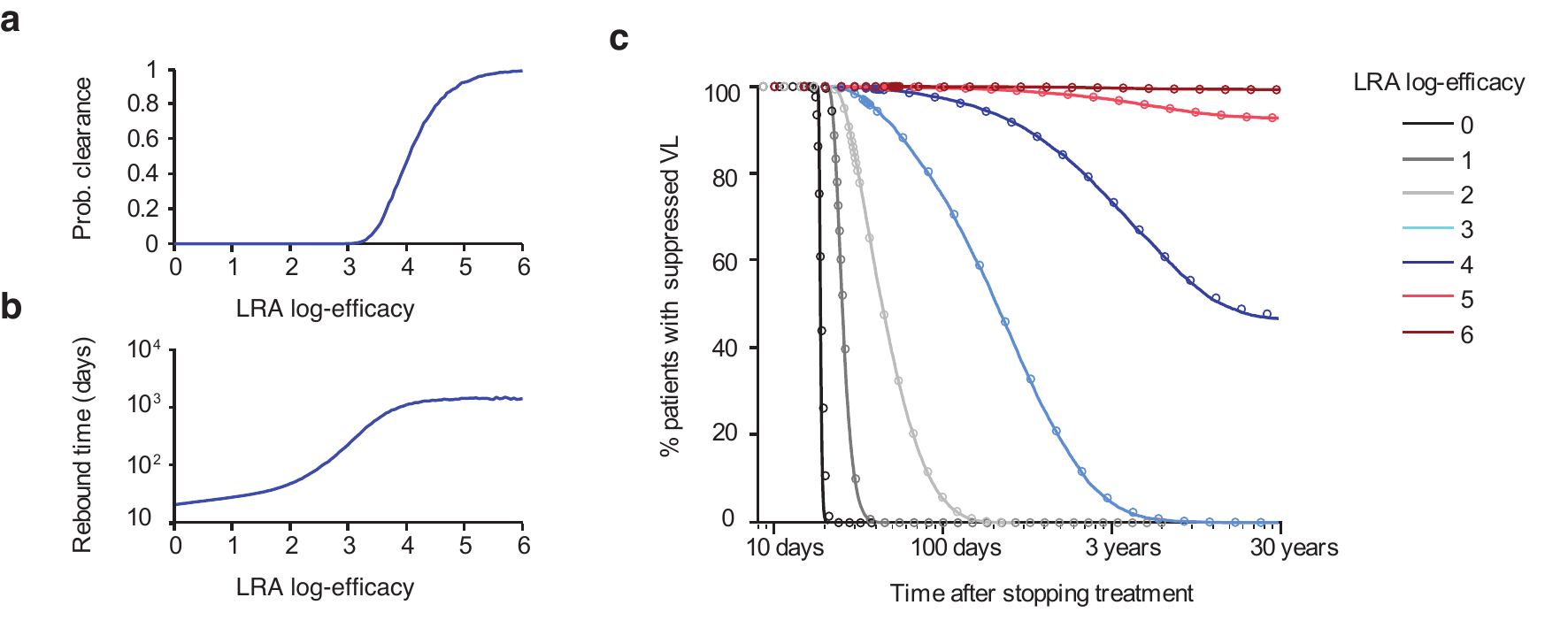}
\caption{Clearance probabilities and rebound times following LRA therapy predicted from model using point estimates for the parameters (Table~\ref{tab:params}). ``LRA log-efficacy'' is the number of orders of magnitude by which the latent reservoir size is reduced following LRA therapy ($-\log_{10}(q)$). a) Probability that the LR is cleared by LRA. Clearance occurs if all cells in the LR die before a reactivating lineage leads to viral rebound. b) Median viral rebound times (logarithmic scale), among patients who do not clear the infection. c) Survival curves (Kaplan-Meier plots) show the percentage of patients who have not yet experienced viral rebound, plotted as a function of the time (logarithmic scale) after treatment interruption. Solid lines represent simulations, and circles represent approximations from the branching process calculation. All simulations included $10^4$ to $10^5$ patients with identical viral dynamic parameter values.}\label{fig:time_rebound}
\end{center}
\end{figure}

The initial conditions for the dynamic model depend on the number of latently infected cells left after LRA therapy. LRA efficacy is defined by the fraction $q$ of the LR that remains following treatment. The model tracks each latent and active cell to determine whether viral rebound occurs, and if so, how long it takes. Importantly, no single activated cell is guaranteed to re-establish the infection, as it may die prior to infecting other cells. Even if it does infect others, those cells likewise may die prior to completing further infection. This possibility is a general property of stochastic models, and the specific value for the establishment probability depends on the rates at which infection and death events occur. Our goal is to calculate the probability that at least one of the infected cells remaining after therapy escapes extinction and causes viral rebound, and if so, how long it takes. If all cells die, then rebound never occurs and a cure is achieved. As the model only describes events after completion of LRA therapy, our results are independent of the therapy protocol or mechanism of action. 
Using both stochastic simulations and theoretical analysis of this model, we find that the probability and timing of rebound relies on four key parameters: the decay rate of the LR in the absence of viral replication ($\delta$), the rate at which the LR produces actively infected cells ($A$), the probability that any one activated cell will produce a rebounding infection before its lineage dies ($P_{Est}$), and the net growth rate of the infection once restarted ($r$). Estimates of these four parameters are provided in Table~\ref{tab:params} and Supplementary Fig.1. After therapy, the rate at which the LR produces actively infected cells is reduced to $q A$. The probability that an individual successfully clears the infection is:
\begin{equation}
P_{Clr}(q) \approx e^{-q A P_{Est} / \delta}.
\label{clearance_prob_maintext}
\end{equation}
The expression $q A P_{Est} / \delta$ approximates the expected number of fated-to-establish cells that will ever exit from the LR, explaining the Poisson form of this expression. In the Supplementary Methods, we provide the full derivation, as well as a formula (Eq. S8) for the probability that rebound occurs a given number of days following treatment interruption (a function of $\delta$, $A$, $P_{Est}$, $r$, and efficacy $q$). Of note, the initial size of the reservoir itself is not included among these parameters:  while it factors into both $A$ (the product of the pre-LRA reservoir size and the per-cell activation rate), and $q$ (the ratio of post-LRA to pre-LRA reservoir size), it does not independently influence outcomes.  Both of these formulas provide an excellent match to explicit simulation of the model (Fig.~\ref{fig:time_rebound}). The key assumption required for the analysis is that $r$ greatly exceeds $\delta$; since viral doubling times during rebound are measured on the order of a few days, while LR decay is measured on the order of many months or years, this assumption is expected to hold. Likelihood-based inference can therefore proceed by efficient computation of rebound probabilities (using equation S8), rather than by time-consuming stochastic simulation.

\begin{table*}[htb!]
\footnotesize
\centering
\begin{tabular}{@{} llp{2.7cm}llll @{}}
\toprule
Parameter & Symbol & Estimation Method & Source & Best Estimate & Distribution\\
\midrule
LR decay rate & $\delta$ & \parbox{2.7cm}{\centering Long-term ART $\left(\delta=\ln(2)/\tau_{1/2}\right)$} & ~\cite{siliciano_long-term_2003, archin_measuring_2014} & $5.2\e{-4}$ d\inv & $\delta \sim \mathcal{N}( 5.2 ,1.6)\e{-4}$ d\inv \\
\rule{0pt}{4.5ex}LR exit rate & $A$ & \multirow{2}{*}{\parbox{2.7cm}{\centering Viral rebound after ART interruption}} & \multirow{2}{*}{~\cite{luo_hiv_2012, ruiz_structured_2000}}  & 57 cells d\inv & $\log_{10}(A) \sim \mathcal{N} (1.76,1.0)$\\
Growth rate & $r$ &  &  & 0.4 d\inv & $\log_{10}(r) \sim  \mathcal{N}(-0.40,0.19)$  \\
\rule{0pt}{4.5ex}Establishment probability & $P_{Est}$ & \parbox{2.7cm}{\centering Population genetic modeling} & ~\cite{pennings_standing_2012, Pennings:2014ib} & 0.069 & \parbox{4cm}{(composite distribution;\newline see Methods)} \\
\bottomrule
\end{tabular}
\caption{Estimated values for the key parameters of the stochastic viral dynamics model \\
{\footnotesize Notation $X \sim \mathcal{N}(\mu,\sigma)$ means that $X$ is a random variable drawn from a normal distribution with mean $\mu$ and standard deviation $\sigma$.}
}
\label{tab:params}
\end{table*}

Outcomes depend only on the four parameters above even in more complex models of viral dynamics including other features of T cell biology and the HIV lifecycle (Supplementary Methods). Alternate models studied include explicit tracking of free virus with varying burst sizes, an ``eclipse phase'' during which an infected cell produces no virus, proliferation of cells upon reactivation, maintenance of the LR by homeostatic proliferation, and either a constant or Poisson-distributed number of infected cells produced by each cell (Supplementary Figs. 2-7). If proliferation of latently infected cells is subject to high variability, e.g., by ``bursts'' of proliferation, then rebound time and cure probability increase slightly beyond the predictions of the basic model (Supplementary Figs. 6 and 7). No other modification to the model altered outcomes. Outcomes of LRA therapy therefore are likely to be insensitive to details of the viral lifecycle; accordingly, few parameters must be estimated to predict outcomes. 

\subsection*{Predicted prospects for eradicating infection or delaying time to rebound}
Using best estimates of parameters derived from previously reported data (Table~\ref{tab:params}), we can explore the likely outcomes of interventions that reduce the latent reservoir. The best outcome of LRA therapy, short of complete and immediate eradication, is that so few latently infected cells survive that none reactivate and start a resurgent infection during the patient's lifespan. In this case, LRA has essentially cleared the infection and a cure is achieved. We simulated the model to predict the relationship between LRA efficacy and clearance (Fig.~\ref{fig:time_rebound}a). We find that the reservoir must be reduced $10{,}000$-fold before half of patients are predicted to clear the infection. 

If LRA therapy fails to clear the infection, the next-best outcome is extension of the time until rebound, defined as plasma HIV-1 RNA $\geq$ 200 c ml$^{-1}$. We computed the relationship between LRA efficacy and median time until rebound among patients who do not clear the infection (Fig.~\ref{fig:time_rebound}b). Roughly a 2{,}000-fold reduction in the reservoir size is needed for median rebound times of 1 year. Only modest ($\sim 2$-fold) increases in median rebound time are predicted for up to 100-fold reductions in LR size. In this range, the rebound time is independent of latent cell lifespan (decay rate $\delta$) and is driven mainly by the reactivation rate ($A$) and the infection growth rate ($r$). The curve inflects upward (on a log scale) at $\approx$100-fold reduction and eventually reaches a ceiling as clearance of the infection becomes the dominant outcome. The upward inflection results from a change in the forces governing viral dynamics. If the reservoir is large (little reduction), then cells activate frequently, and the dominant component of rebound time is the time that it takes for virus from the many available activated cells to grow exponentially to rebound levels; the system is in a \emph{growth-limited regime}. If the reservoir is small (large reduction), the dominant component is instead the expected waiting time until activation of the first cell fated to establish a rebounding lineage; the system is in an \emph{activation-limited regime}. Since waiting time is roughly exponentially distributed, times to rebound in this regime can vary widely among patients on the same therapy, even with identical values of the underlying parameters.

Survival curves, plotting the fraction of simulated patients maintaining virologic suppression over time, demonstrate the extreme interpatient variability and long follow-up times required for LRA therapy (Fig.~\ref{fig:time_rebound}c). For less than 100-fold reductions in LR size, simulated patients uniformly rebound within a few months, since rebound dynamics are not in the activation-limited regime. If therapy decreases LR size $1{,}000$-fold, then $\sim$55\% of patients are predicted to delay rebound for at least six months. However, of these patients, 47\% suffer rebound in the following six months. Higher reservoir reductions lead to clearance in many patients. In others, rebound may still occur after years of apparent cure, posing a challenge for patient management.

Earlier work suggested a shorter reservoir half-life of 6 months ~\cite{zhang_quantifying_1999}, indicating that dramatic decreases in LR size would occur after 5 or more years of suppressive ART even in the absence of LRA therapy.  We consider the prospects for HIV eradication or long treatment interruptions with this faster reservoir decay rate (Fig.~\ref{fig:time_rebound_vary}b). In this optimistic scenario, only 1{,}500-fold reductions are needed for half of patients to clear the LR, and rebound becomes highly unlikely after a few years. Alternatively, in a worst-case scenario where latent cell death is perfectly balanced by homeostatic proliferation such that the reservoir does not decay at all ($\delta = 0$), much higher efficacies are needed to achieve beneficial patient outcomes (Fig.~\ref{fig:time_rebound_vary}c).  

\subsection*{Setting treatment goals with uncertainty considerations}

\begin{SCfigure*}[0.8][htb!]
 \includegraphics[width=1.1\columnwidth]{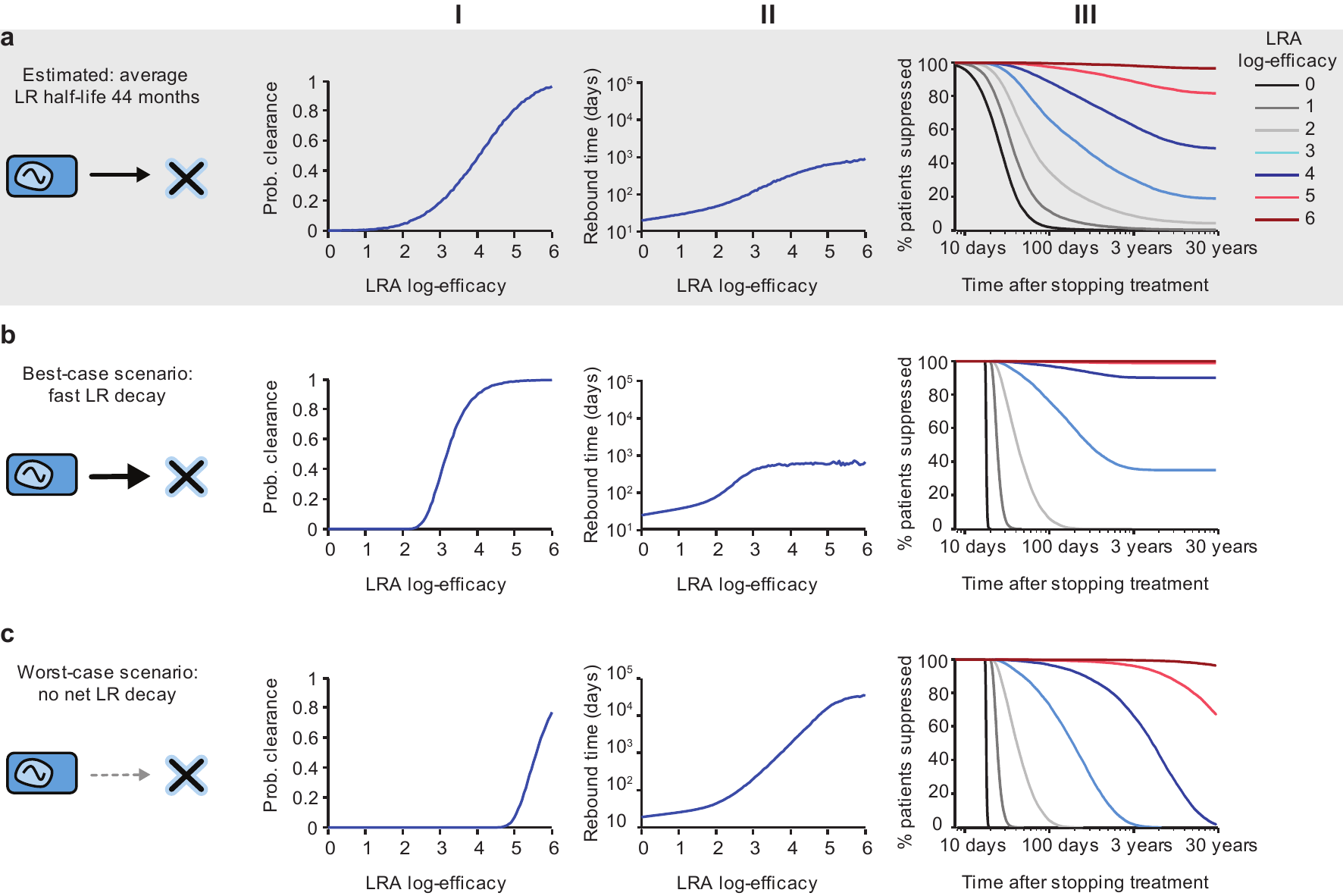}
\caption{Predicted LRA therapy outcomes, accounting for uncertainty in patient parameter values. 
a) Full uncertainty analysis where all viral dynamics parameters are sampled for each patient from the distributions provided in Table~\ref{tab:params}.
b) A best-case scenario where the reservoir half-life is only 6 months ($\delta = 3.8\e{-3}$ $d^{-1}$). All patients have the same underlying viral dynamic parameters, otherwise given by the point estimates in Table~\ref{tab:params}. 
c) A worst-case scenario where the reservoir does not decay because cell death is balanced by homeostatic proliferation ($\delta = 0$).
I) Probability that the LR is cleared by LRA. Clearance occurs if all cells in the LR die before a reactivating lineage leads to viral rebound. ``LRA log-efficacy'' is the number of orders of magnitude by which the latent reservoir size is reduced following LRA therapy ($-\log_{10}(q)$).
II) Median viral rebound times (logarithmic scale), among patients who do not clear the infection. 
III) Survival curves (Kaplan-Meier plots) show the percentage of patients who have not yet experienced viral rebound, plotted as a function of the time (logarithmic scale) after treatment interruption. All simulations included $10^4$ to $10^5$ patients.} 
\label{fig:time_rebound_vary}
\end{SCfigure*}

We conducted a full uncertainty analysis of the model, by simultaneously varying all parameters over their entire ranges (Table~\ref{tab:params}, Supplementary Fig. 1). For each simulated patient, values for the three parameters $\delta$, $A$, and $r$ were sampled independently from their respective distributions, while $P_{Est}$ was sampled from a conditional distribution that depends on $r$ (see Methods). Results for this simulated cohort are similar to those for the point estimates, with greater interpatient variation in outcomes (Fig.~\ref{fig:time_rebound_vary}a). This variation makes the survival curves less steep: cure is slightly more likely at low efficacy, but slightly less likely at high efficacy. As expected from Equation~\eqref{clearance_prob_maintext}, cure is more likely for patients with lower $A$ or $P_{Est}$ values and higher $\delta$ values. If therapy provides only 10--100-fold LR reductions, a subset of patients may delay rebound for several months.

Using these cohort-level predictions, we can set efficacy goals for the reservoir reduction needed to achieve a particular likelihood of a desired patient outcome. Fig.~\ref{fig:qcrit} provides the target LRA efficacies for which 50\% of patients are predicted to remain rebound-free for a specified interruption time. Reductions of under 10-fold afford patients only a few weeks to a month off treatment without rebound. For one-year interruptions, a 1{,}000--3{,}000-fold reduction is needed. To achieve the goal of eradication (cure) a 4-log reduction is required. This value increases to 4.8 logs to cure 75\% of patients, and to 5.8 logs for 95\% of patients.

\subsection*{Model applications and comparison to data}
\begin{figure}[htb!]
\begin{center}
\includegraphics[]{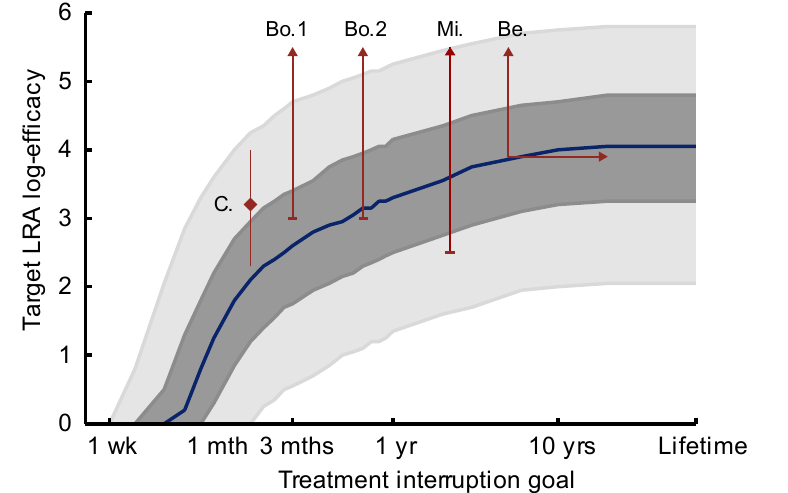}
\caption{Efficacies required for successful LRA therapy. The target LRA log-efficacy is the treatment level (in terms of log-reduction in latent reservoir size) for which at least 50\% of patients still have suppressed viral load after a given treatment interruption length (blue line). Shaded ranges show the results for the middle 50\% (dark gray) and 90\% (light gray) of patients. ``Lifetime'' means the LR is cleared.  Annotations on the curve represent data points for case studies describing large reservoir reductions and observing rebound times after ART interruption. 
From left to right, they represent a case of early ART initiation in an adult (the ``Chun patient'' (C.) ~\cite{chun_rebound_2010}), two cases of hematopoietic stem cell transplant with wild-type donor cells (the two Boston patients (Bo.1 and Bo.2) ~\cite{henrich_antiretroviral-free_2014}), a case of early ART initiation in an infant (the Mississippi baby (Mi.) ~\cite{persaud_very_2014}, assuming, as recently reported, rebound after 27 months), and a case of hematopoietic stem cell transplant with $\Delta$32 CCR5 donor cells (the Berlin patient (Be.) ~\cite{hutter_long-term_2009,yukl_challenges_2013}). 
For the Chun patient, the annotations represent the maximum likelihood estimate for LR reduction (diamond), as well as 95\% confidence intervals (vertical bar). For the Boston and Berlin patients, vertical arrows indicate that only a lower bound on treatment efficacy is known (LR size was below the detection limit) and that the true value may extend further in the direction shown. For the Berlin patient, the horizontal arrow indicates that rebound time is at least five years (rebound has not yet occurred). 
}
\label{fig:qcrit}
\end{center}
\end{figure}

Current ability to test the model against clinical data is limited both by the dynamic range of assays measuring LR size and by the low efficacy of investigational LRA treatments. Yet we can compare our predictions to results observed for non-LRA-based interventions that lead to smaller LR size and prolonged treatment interruptions (Fig.~\ref{fig:qcrit}). A 2010 study of early ART initiators who eventually underwent treatment interruption found a single patient with LR size approximately 1{,}500-fold lower than a typical patient (0.0064 infectious units per million resting \cdfour T cells, versus an average of 1 per million) in whom rebound was delayed until 50 days off treatment~\cite{chun_rebound_2010}. The well-known `Berlin patient'~\cite{hutter_long-term_2009} has remained off treatment following a stem-cell transplant since 2008, and a comprehensive analysis of his viral reservoirs found HIV DNA levels \emph{at least} 7{,}500-fold lower than typical patients in the most sensitive assay ~\cite{yukl_challenges_2013}. The two recently reported `Boston patients' also interrupted treatment, following transplants that caused at least a 3 to 4 $\log$ decrease in viral reservoirs~\cite{henrich_antiretroviral-free_2014}; they have since both rebounded, at approximately 3 and 8 months post-interruption. In the case of the `Mississippi baby', infection was discovered and treated within 30 hours of birth, and ART continued until interruption at around 18 months. Virus remained undetectable for 27 months, when viral rebound occurred, assuming the accuracy of widely reported claims ~\cite{ledford_hiv_2014}.  At the time of treatment cessation, the LR size was likely at least 300-fold lower than that of a typical adult (based on less than 0.017 infectious units per million resting CD4+ T cells at age 30 months ~\cite{persaud_very_2014}, and scaled on a weight basis relative to adults). These few available cases demonstrate that our model is not inconsistent with current knowledge. When survival curves for larger cohorts become available, Bayesian methods can be used to update estimates in Table~\ref{tab:params} and reduce uncertainty of future predictions.

\section*{Discussion}
Our model is the first to quantify the required efficacy of latency-reversing agents for HIV-1 and set goals for therapy.  For a wide range of parameters, we find that therapies must reduce the LR by at least two orders of magnitude to meaningfully increase the time to rebound after ART interruption (upward inflection in Figs.~\ref{fig:time_rebound}b,~\ref{fig:time_rebound_vary}II), and that reductions of approximately four orders of magnitude are needed for half of patients to clear the infection (Figs.~\ref{fig:time_rebound_vary}a,~\ref{fig:qcrit}). Standard deviations in rebound times of many months are expected, owing to substantial variation in reactivation times after effective LRA therapy brings the infection to an activation-limited regime. While the efficacy required for these beneficial outcomes is likely beyond the reach of current drugs, our results permit some optimism: we show for the first time that reactivation of all cells in the reservoir is not necessary for cessation of ART. This is because some cells in the LR will die before reactivating or, following activation, will fail to produce a chain of infection events leading to rebound. On a more cautionary note, the wide distribution in reactivation times necessitates careful monitoring of patients, as rebound may occur even after long periods of viral suppression.

Even without any reservoir reduction, variation in infection parameters and chance activation together predict delays in rebound of at least two months in a small minority of patients (Fig.~\ref{fig:qcrit}), consistent with ART interruption trials such as SPARTAC ~\cite{Stohr:2013hb}. More detailed (and possibly more speculative) models including specific immune responses may be needed to explain multi-year post-treatment control, such as found in the VISCONTI cohort ~\cite{saez-cirion_post-treatment_2013}.

Our analysis characterizing the required efficacy of LRA therapy does not rely on the specific mechanism of action of these drugs, only the amount by which they reduce the reservoir. We have assumed that, after ART/LRA therapy ends, cell activation and death rates return to baseline.  We have also assumed that the reservoir is a homogeneous population with constant activation and death rates. The presence of reservoir compartments with different levels of LRA penetration does not alter our results, as they are stated in terms of total reservoir reduction. If, however, these compartments vary in activation or death rates  ~\cite{Buzon:2014hb}, or if dynamics of activated cells depends on their source compartment, then our model may need to be modified. Moreover if spatial population structure affects viral replication, viral dynamics above the detection limit (from which we estimated parameters $r$ and $A$) may not correspond straightforwardly to the infection/death rates in early infection, due to local limitations in target cell density ~\cite{strain_spatiotemporal_2002}. Spatial restrictions on viral transmission may be particularly important in densely packed lymphoid tissue ~\cite{cardozo_spatial_2014}. In the absence of clear understanding of multiple compartments constituting the LR, we have considered the simplest scenario which may fit future LRA therapy outcomes.

Throughout this paper, we assume that combination ART is sufficiently effective so that viral replication alone cannot sustain the infection after all latent virus is cleared. Studies of treatment intensification ~\cite{gandhi_effect_2010,dinoso_treatment_2009}, of viral evolution during ART ~\cite{kieffer_genotypic_2004,joos_hiv_2008}, and of \emph{in vitro} antiviral efficacy ~\cite{shen_dose-response_2008,jilek_quantitative_2012} all support this assumption. Moreover, HIV persistence is widely believed to result solely from the long lifespan or proliferative ability of latently infected cells ~\cite{siliciano_long-term_2003,chomont_hiv_2009}. If this assumption is violated, e.g., by the presence of long-lived drug-protected compartments ~\cite{luo_modelling_2013,cardozo_spatial_2014, fletcher_persistent_2014}, then any curative strategy predicated solely on latency reversal would be futile.

Our model also highlights the importance of measuring specific parameters describing latency and infection dynamics. Despite the field's focus on measuring latent reservoir size with increasing accuracy ~\cite{ho_replication-competent_2013}, our results suggest that the \emph{rate} at which latently infected cells activate --- and the fraction of these that are expected to establish a rebounding infection --- are more predictive of LRA outcomes. Among all parameters that determine outcome, the establishment probability is least understood, as it cannot be measured from viral load dynamics above the limit of detection. Simply because an integrated provirus is replication-competent and transcriptionally active does not mean that it \emph{will} initiate a growing infection: as with all population dynamics, chance events dominate early stages of infection growth ~\cite{pearson_stochastic_2011, hofacre_early_2012}. HIV-1 transcription is itself a stochastic process, governed by fluctuating concentrations of early gene products ~\cite{Singh:2009bg}. Sensitive assays of viral outgrowth may pave the way toward understanding the importance of these chance events to early infection; for instance, fluorescent imaging studies of adenovirus have shown that a large majority of \emph{in vitro} infections seeded by single productively infected cells die out early, before rapid growth and plaque formation can occur~\cite{hofacre_early_2012}. Similar experiments with HIV are underway in our laboratory and may help refine parameter estimates. Keeping other parameters constant, assuming a worst-case (highest) value for the establishment probability raises the reservoir reductions required for cure or a desired extended rebound time by 0.8 logs. Regardless of the exact probability, the stochastic nature of HIV-1 activation and infection dynamics implies that even similarly situated patients may experience divergent responses to LRA.

The model can also advise aspects of trial design for LRAs. Survival curves computed from equation S8 can be used to predict the probability that a patient is cured, given that they have been off treatment without rebound for a known period. As frequent viral load testing for years of post-interruption monitoring is not feasible, it may be helpful to choose sampling timepoints based on the expected distribution of rebound times. Trial design is complicated by the fact that LRA treatment efficacy is unknown if post-treatment LR size is below the detection limit. By considering prior knowledge about viral dynamics parameters and the range of possible treatment efficacies, the model may estimate outcomes even in the presence of uncertainty.

To date, laboratory and clinical studies of investigational LRAs have generally found weak potential for reservoir reduction --- up to one log-reduction \emph{in vitro} and less \emph{in vivo} ~\cite{archin_antiretroviral_2010,xing_disulfiram_2011,cillo_quantification_2014}. We predict that much higher efficacy will be required for eradication, which may be achieved by multiple rounds of LRA therapy, a combination of therapies, or development of therapies to which a greater fraction of the LR is susceptible. While we have focused on LRA therapy, our findings also serve to interpret infection eradication or delays in rebound caused by early treatment~\cite{Strain:2005hu,saez-cirion_post-treatment_2013,persaud_absence_2013} or stem cell transplantation~\cite{hutter_long-term_2009,henrich_antiretroviral-free_2014}, both of which also reduce the latent reservoir. In both of these cases, however, additional immunological dynamics likely play a major role and will need to be incorporated into future models. We believe that these modeling efforts will provide a quantitative framework for interpreting clinical trials of any reservoir-reduction strategy.

\section*{Methods}
\subsection*{Basic stochastic model} The basic model of reservoir dynamics and rebound tracks two cell types: productively infected activated \cdfour T cells, and latently infected resting \cdfour T cells. The model can be described formally as a two-type branching process, in which four types of events can occur (Fig.~\ref{fig:modeldiagram}):
\begin{equation}
\begin{split}
Z &\to Y \mbox{... rate constant: } a\\
Z &\to \emptyset \mbox{... rate constant: } d_z\\
Y &\to cY \mbox{... rate constant: } b \times p_{\lambda}(c)\\
Y &\to \emptyset \mbox{... rate constant: } d\\
\end{split}
\end{equation}
In this notation $Y$ and $Z$ represent individual actively or latently infected cells, respectively, $\emptyset$ represents no cells, and the arrows represent one type of cell becoming the other type. A latently infected cell can either activate (at rate $a$) or die (at rate $d_z$). An actively infected cell can either die (at rate $d$) or produce a collection of virions (at rate $b$) that results in the infection of $c$ other cells, where $c$ is a Poisson-distributed random variable with parameter $\lambda$, $p_{\lambda}(c) = \left(\exp(-\lambda) \lambda^c\right)/(c!)$. After an infection event, the original cell dies. 

Each event occur independently within a large, constant target cell population. As the model does not include limitations on viral growth, it describes only the initial stages of viral rebound. Since clinical rebound thresholds (plasma HIV RNA $>$ 50 -- 200 c ml$^{-1}$) are well below typical setpoints ($10^4$ -- $10^6$ c ml$^{-1}$), this model suffices to analyze rebound following LRA therapy and ART interruption. We do not explicitly track free virus, but assume it to be proportional to the number of infected cells. This assumption is valid because rates governing production of and clearance of free virus greatly exceed other rates, allowing a separation of time scales. As we are not interested in blips or other intraday viral dynamics, this assumption does not influence our results. A method for calculating the proportionality between free virus and infected cells is provided in the Supplementary Methods.

The growth rate of the infection is $r=b(\lambda-1)-d$. The total death rate of infected cells is $d_y = b+d$, and the basic reproductive ratio (mean offspring number for a single infected cell) is $R_0=b \lambda/(b + d)$. The establishment probability $P_{Est}$ is the solution to $R_0 \left(1-e^{-\lambda P_{Est}}\right) - \lambda P_{Est} = 0$. The total LR decay rate in the absence of viral replication is $\Delta = a + d_z$. If there are $\mathcal{Z}$ cells in the latent reservoir, then the number of cells reactivating per day is $A = \mathcal{Z} a$. 

Analysis of the model to determine the four key parameters ($\delta$, $A$, $r$, $P_{Est}$) and rapidly compute survival curves is provided in the Supplementary Methods. A script for computation of survival curves is also provided at http://www.danielrosenbloom.com/reboundtimes.

\subsection*{Parameter estimation}The half-life of latently infected cells has been estimated to be approximately $\tau_{1/2} = 44$ months ~\cite{siliciano_long-term_2003,archin_measuring_2014}. The resulting value of $\delta = \ln(2)/\tau_{1/2}$ is centered at $5.2\e{-4}$ d$^{-1}$, and we construct a distribution of values based on ref.~\cite{siliciano_long-term_2003} . This value represents the \emph{net} rate of LR decay during suppressive therapy, considering activation, death, homeostatic proliferation, and (presumably rare) events where activated \cdfour T cells re-enter a memory state. The net infection growth rate $r$ describes the rate of exponential increase in viral load once infection has been reseeded. The LR reactivation rate $A$ is the number of cells exiting the LR per day, before reservoir-reducing therapy. $A$ and $r$ were jointly estimated from the dynamics of viral load during treatment interruption trials in which there was no additional reservoir-reducing intervention~\cite{luo_hiv_2012, ruiz_structured_2000}; in particular, infection growth immediately following rebound is sensitive to $r$, while the time to rebound is sensitive to $A$. In the absence of reservoir reduction, observed rebound dynamics are insensitive to $P_{Est}$, and so this parameter was instead estimated from population genetic models ~\cite{pennings_standing_2012, Pennings:2014ib} that relate observed rates of selective sweeps and emergence of drug resistance to variance in the viral offspring distribution (see Supplementary Methods).

\subsection*{Simulation of the model} We use the Gillespie algorithm to track the number of latently and actively infected cells in a continuous time stochastic process. The initial number of latent cells is $\mathcal{Z}(0) \sim Binomial(N_{LR},q)$, where $N_{LR}$ is the pre-treatment reservoir size and $q$ is the efficacy of LRA treatment (fraction of cells remaining). The initial number of actively infected cells $\mathcal{Y}(0)$ is then chosen from a Poisson distribution with parameter $a \mathcal{Z}(0) /d_y$ (corresponding to the immigration-death equilibrium of the branching process). The simulation proceeds until the number of actively infected cells reaches the threshold for clinical detection given by a viral load of 200 c ml$^{-1}$ (equivalent to $\mathcal{Y}=3\e{5}$ cells total) or until no active or latent cells remain. Because stochastic effects are important only for small $\mathcal{Y}$, we switch to faster deterministic numerical integration when $\mathcal{Y}$ reaches a level where extinction probability is very low ($<10^{-4}$). For each $q$ value we perform $10^4$ to $10^5$ simulations.

Simulations are seeded with values of the key parameters ($\delta$, $A$, $r$, $P_{Est}$), which may be either the point estimates or random numbers sampled from the distributions in Table~\ref{tab:params}. We then back out values of the model-specific parameters that are consistent with the sampled key parameters. In general, we use a pre-therapy LR size of $N_{LR} = 10^6 $ cells to get $a = A/N_{LR}$. We then have $d_z = \delta - a$.
As detailed in the Supplementary Methods, sampling $P_{Est}$ requires first sampling the variance-to-mean ratio of the viral offspring distribution ($\rho$). Then using $r$ and $\rho$ along with $d_y = d+b$ = 1 day\inv, we can get $\lambda$, $b$, $d$, and $P_{Est}$. Consistent with our generating function analysis, we find that the specific values assumed for $N_{LR}$ and $d_y$ do not influence the results. For simulating other models, any other parameter assumptions are listed in the corresponding supplementary figure captions.

\section*{Acknowledgements}
We thank Y.-C. Ho, S. A. Rabi, L. Shan, and G. Laird for insightful discussions and for sharing data, and we thank A. Perelson for helpful comments on an earlier version of this manuscript. This work was supported by the Martin Delaney CARE and DARE Collaboratories (NIH grants AI096113 and 1U19AI096109), by an ARCHE Grant from the American Foundation for AIDS Research (amFAR 108165-50-RGRL), by the Johns Hopkins Center for AIDS Research, and by the Howard Hughes Medical Institute. MAN was supported by the John Templeton Foundation. FF was funded by a European Research Council Advanced Grant (PBDR 268540). ALH and DISR were supported by a Bill \& Melinda Gates Foundation Grand Challenges Explorations Grant (OPP1044503).

\bibliography{main}
\bibliographystyle{nature}

\end{document}